\documentclass[%
reprint,
superscriptaddress,
nofootinbib,
 amsmath,amssymb,
 aps,
]{revtex4-2}

\usepackage{graphicx}
\usepackage{dcolumn}
\usepackage{bm}
\usepackage{braket}
\usepackage{upgreek}
\usepackage{xcolor}
\usepackage{afterpage}
\usepackage{relsize}
\usepackage{multirow}
\usepackage{float}
\usepackage{physics}
\usepackage{makecell}
\usepackage{placeins}
\usepackage{amsthm}
\usepackage[colorlinks=true,urlcolor=blueprl,citecolor=blueprl,linkcolor=blueprl]{hyperref}
\usepackage{lineno}
\usepackage[symbol]{footmisc}
\usepackage{hhline}
\usepackage[normalem]{ulem}

\newcommand{\ozlem}[1]{\textcolor{black}{#1}}
\newcommand{\ozlemblue}[1]{\textcolor{black}{#1}}
\newcommand{\ozlemREV}[1]{\textcolor{black}{#1}}

\newcolumntype{P}[1]{>{\centering\arraybackslash}p{#1}}

\usepackage{hyperref}
\let\svthefootnote\thefootnote
\newcommand\freefootnote[1]{%
  \let\thefootnote\relax%
  \footnotetext{#1}%
  \let\thefootnote\svthefootnote%
}

\definecolor{blueprl}{RGB}{46,48,146}

\def\UQA{Centre for Quantum Computation and Communication Technology, School of Mathematics and Physics, University of Queensland, St Lucia, QLD 4072, Australia}
\def\ASTAR{A*STAR Quantum Innovation Centre (Q.InC), Agency~for~Science,~Technology~and~Research~(A*STAR), 2 Fusionopolis Way, Innovis \#08-03, Singapore 138634, Republic of Singapore}
\def\ANU{Centre of Excellence for Quantum Computation and Communication Technology, The~Department~of~Quantum Science and Technology, Research School of Physics and Engineering, The Australian National University, Canberra, Australian Capital Territory, Australia}
\def\QCTRL{Present address: Q-CTRL, Sydney, New South Wales, Australia}

\begin{document}


\title{A unified optical platform for non-Gaussian and fault-tolerant Gottesman-Kitaev-Preskill states}
\author{\"{O}zlem Erk{\i}l{\i}\c{c}$^{*}$}
\freefootnote{$^*$ \href{ozlemerkilic1995@gmail.com}{ozlemerkilic1995@gmail.com}}
\affiliation{\UQA}
\affiliation{\ANU}

\author{Aritra Das}
\affiliation{\ANU}

\author{Biveen Shajilal}
\affiliation{\ASTAR}
\affiliation{\QCTRL}

\author{Ping Koy Lam}
\affiliation{\ASTAR}

\author{Timothy C. Ralph}
\affiliation{\UQA}

\author{Syed M. Assad}
\affiliation{\ASTAR}

\date{\today}
             
\maketitle

\section*{Abstract}
Quantum technologies, encompassing communication, computation, and metrology, rely on the generation and control of non-Gaussian states of light. These states enable secure quantum communication, fault-tolerant quantum computation, and precision sensing beyond classical limits, yet their practical realisation remains a major challenge due to reliance on high-photon-number Fock states \ozlemREV{or} strong non-linearities. Here we introduce a unified optical framework that removes this constraint, using only Gaussian inputs, optical parametric amplification, and heralded photon detection. Within a single architecture, we demonstrate the generation of photon-added squeezed states with near unit fidelity, cubic-phase-like states with strong non-linearities and fidelities above $98.5\%$, and squeezed-cat states exceeding $99\%$ fidelity that can be iteratively bred into GKP grid states surpassing the 9.75 dB fault-tolerance threshold. Operating entirely below 3 dB of input squeezing, the approach provides a scalable, experimentally accessible platform that unites the state resources required for quantum communication, metrology, and computation within one coherent optical framework.

\section{\label{sec:introduction}Introduction}
Quantum technologies~\cite{o2009photonic, wang2020integrated} encompass a diverse range of platforms, including quantum metrology~\cite{leibfried2004toward,kacprowicz2010experimental, krischek2011useful, toth2014quantum, daryanoosh2018experimental, guo2020distributed, marciniak2022optimal, conlon2023approaching, conlon2023discriminating, conlon2024verifying, waslh2025all}, quantum communications~\cite{ch1984quantum, bouwmeester1997experimental, gisin2007quantum, ursin2007entanglement, kimble2008quantum, pirandola2015advances, valivarthi2016quantum, pirandola2020advances, erkilicc2023surpassing, liu2023experimental, erkilic2024capacity, erkilic2025software}, and quantum computing~\cite{deutsch1985quantum, knill2001scheme, kok2007linear, preskill2018quantum, flamini2018photonic, zhong2020quantum, madsen2022quantum}. Although these technologies pursue different goals, they utilise similar classes of quantum states. Depending on the physical platform, quantum information can be encoded in discrete-variable~(DV) systems, where it is carried by binary degrees of freedom such as photon polarisation, or in continuous-variable~(CV) systems, which use the amplitude and phase quadratures of the optical field to encode information. In CV architectures, the fundamental operations and states are generally Gaussian, defined solely by their first and second statistical moments, and are experimentally accessible through linear optics, squeezing, and homodyne detection~\cite{weedbrook2012gaussian}. However, the performance of Gaussian-only platforms is inherently limited, falling short of enabling the ultimate precision scaling~\cite{giovannetti2004quantum, escher2011general, demkowicz2015quantum}, entanglement distillation~\cite{eisert2002distilling} and universal quantum computing~\cite{lloyd1999quantum, knill2001scheme, bartlett2002efficient, mari2012positive}.

Non-Gaussian states and operations, on the other hand, allow these platforms to reach their full potential~\cite{lloyd1999quantum, knill2001scheme, bartlett2002efficient, mari2012positive, erkilic2025enhanced}. One class of such states are the photon added or subtracted Gaussian states which are typically generated via interacting a coherent or squeezed state of light with Fock states on a beamsplitter and performing photon-number–resolving measurements~\cite{zavatta2004quantum, wenger2004non, fiuravsek2005conditional, takahashi2008generation, su2019conversion, ra2020non, asavanant2021wave, endo2023non, endo2024multi, endo2025high}. These states have been shown to enhance the performance across quantum communications and metrology. In quantum key distribution~(QKD), photon-added or photon-subtracted Gaussian states can improve achievable key rates~\cite{huang2013performance, li2016non, zhong2018self, jeng2025entanglement}, while in quantum teleportation they increase the fidelity of transmitted states~\cite{wang2015continuous, arora2025continuous} and in quantum metrology, they lead to more precise phase estimation~\cite{ouyang2016quantum, zhao2025improved} and reduce detection error rates~\cite{zhang2024quantum}. While these states offer clear advantages, extending photon addition beyond two photons~\cite{bimbard2010quantum} poses significant experimental challenges, primarily due to the difficulty of producing and interfering higher-order Fock states with sufficient purity and efficiency.

The cubic phase state is another important non-Gaussian resource, providing the non-linearity required for universal CV quantum computation and non-linear gate operations~\cite{lloyd1999quantum,gottesman2001encoding,bartlett2002efficient,gu2009quantum}. It is typically generated by performing photon-number–resolving measurements on Gaussian input states, where conditioning on detection outcomes induces the desired phase structure~\cite{gottesman2001encoding,yukawa2013emulating,sabapathy2018states}, or via Kerr non-linearities that directly implement the cubic phase shift~\cite{yanagimoto2020engineering}. However, optical non-linearities achieved to date have produced only weak cubic phase states~\cite{marek2011deterministic,yanagimoto2020engineering}. Recent work by Jeng~\textit{et al.}~\cite{jeng2024strong} showed that photon addition to coherent states can yield stronger cubic phase profiles, though this was demonstrated through post-processing of heterodyne data rather than by generating a physical state. Realising such a scheme experimentally would likewise require high-order Fock-state generation, presenting the same practical limitation as photon-added state preparation.

A further class of non-Gaussian resources essential for CV fault-tolerant quantum computing is the Gottesman-Kitaev-Preskill~(GKP) states~\cite{gottesman2001encoding, menicucci2014fault, fukui2018high}. GKP states can be generated using Gaussian Boson Sampling~(GBS)~\cite{zhong2019experimental, deng2023gaussian, liu2025robust}, where an anti-squeezed state is interfered with $N$ squeezed states, and the resulting output modes are measured with $N$ PNRDs~\cite{su2019conversion, sabapathy2019production, tzitrin2020progress, fukui2022efficient}. However, determining the sampling parameters required to generate GKP qubits is a complex numerical problem, and the resulting solutions typically offer limited error-correction capability or low generation rates~\cite{takase2023gottesman}. Nevertheless, GKP states have recently been demonstrated experimentally using GBS platforms, however, the achieved states are not yet at the level required for fault-tolerant operation~\cite{larsen2025integrated}. Alternatively, GKP states can be constructed from cat or squeezed-cat states via cat-breeding protocols~\cite{vasconcelos2010all,weigand2018generating}, in which multiple cat states are interfered and measured using homodyne detection~\cite{konno2024logical}. Early proposals for generating cat or squeezed-cat states relied on interfering a Fock state with vacuum and performing homodyne detection~\cite{ourjoumtsev2006generating}, or alternatively, on using squeezed-vacuum states and conditional photon-number measurements on a beamsplitter to realise photon subtraction~\cite{dakna1997generating, etesse2015experimental, endo2023non}. However, these schemes typically produced states with either low amplitude or limited squeezing or were highly non-deterministic~\cite{ourjoumtsev2006generating, takase2021generation, zhang2022all, takase2022gaussian, podoshvedov2022algorithm, eaton2022measurement}. To address these challenges, \ozlemREV{Ref.~\cite{winnel2024deterministic}} developed a high-fidelity protocol for generating squeezed-cat and approximate GKP states. In their scheme, a Fock state is interfered with multiple vacuum ancillas on $N+1$ beamsplitters, after which squeezing operations are applied and the $N$ ancilla modes are measured using PNRDs. However, the protocol becomes deterministic only in the limit of a large Fock-state input, where the practical generation rate of such Fock states is not accounted for. In practice, the difficulty of producing high-photon-number Fock states beyond the two-photon level remains a major limitation, as their generation success probability rapidly vanishes with increasing photon number. \ozlem{Recently, Fock-state free approaches based on the photon-counting-assisted node-teleportation method~\cite{renault2025end, renault2025squeezing} can also generate squeezed-cat and GKP states, but they require deep teleportation circuits with multiple squeezing and anti-squeezing operations, making the overall implementation comparatively complex.}
\begin{figure}[t]
\includegraphics[scale=0.8]{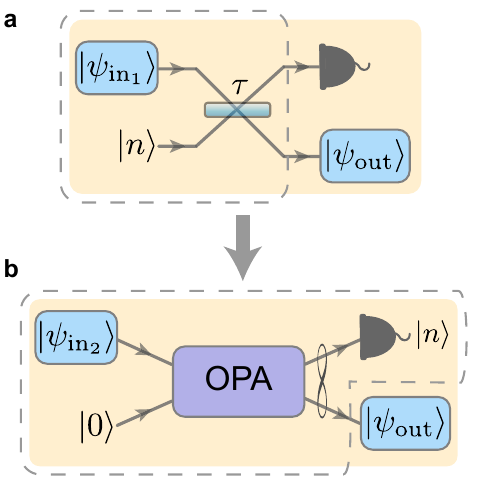}\hspace*{-0.1cm}
\caption{\label{fig:figure1}\textbf{Schematic of the proposed non-Gaussian source.} \textbf{(a)} Generic scheme for generating non-Gaussian states from a Gaussian input $\ket{\psi_{\mathrm{in}_1}}$. The input state is interfered with a Fock state $\ket{n}$ on a beamsplitter with transmissivity, $\tau$, and the desired non-Gaussian output, $\ket{\psi_{\mathrm{out}}}$, is heralded by a PNRD. \textbf{(b)} In the proposed non-Gaussian source, a Gaussian seed state, $\ket{\psi_{\mathrm{in}_2}}$, is injected into an optical parametric amplifier~(OPA), generating two correlated output modes. PNRD on one mode heralds the generation of a non-Gaussian state in the other.
}
\end{figure}

Nearly all non-Gaussian state-generation schemes rely on Fock states as a non-Gaussian resource, which fundamentally limits their scalability. Here, we propose an alternative approach that removes the need for Fock-state inputs altogether. Our scheme, based solely on Gaussian states and operations combined with PNRDs, enables the generation of photon-added Gaussian states, cubic-phase states, and squeezed-cat states required for various quantum operations with high fidelities. The cubic-phase states produced in this framework exhibit strong non-linearities, making them suitable for implementing high-fidelity non-linear gates. Furthermore, the squeezed-cat states generated through this approach can be used to realise GKP states that exceed the fault-tolerance threshold for quantum computing, requiring only low levels of squeezing and single-photon detection. This framework establishes a unified and experimentally practical platform for non-Gaussian state generation, providing a common foundation for quantum metrology, communications, and computation.

\section{\label{sec:non_gaussian_source}The Non-Gaussian Source}
Figure~\ref{fig:figure1}(a) illustrates a generic method for producing non-Gaussian states from Gaussian inputs, where a Gaussian state $\ket{\psi_{\mathrm{in}}}$ is mixed with a Fock state $\ket{n}$ on a beamsplitter, and a PNRD on one mode heralds the non-Gaussian output. In contrast, our proposed non-Gaussian source, shown in Fig.~\ref{fig:figure1}(b), eliminates the need for Fock-state generation by seeding a Gaussian state into an optical parametric amplifier (OPA). \ozlemREV{A related OPA-based scheme was previously explored in Ref.~\cite{shringarpure2019generating} by seeding the signal mode with a coherent state and the idler with a heralded single-photon Fock state, followed by single-photon detection at the output to generate photon-added states and some other non-Gaussian states. However, this approach still requires a pre-generated single-photon Fock state as the seed in the idler and is restricted to single-photon heralding. In contrast, our architecture operates entirely with Gaussian inputs, removes the need for any Fock-state resources, and supports multi-photon heralding for generating a much broader class of non-Gaussian states.} 

The OPA produces two correlated output modes, and PNRD on one mode naturally heralds a non-Gaussian state in the other. The OPA in our setup can be described by the two-mode squeezing unitary
\begin{figure*}[htbp]
\includegraphics[scale=0.36]{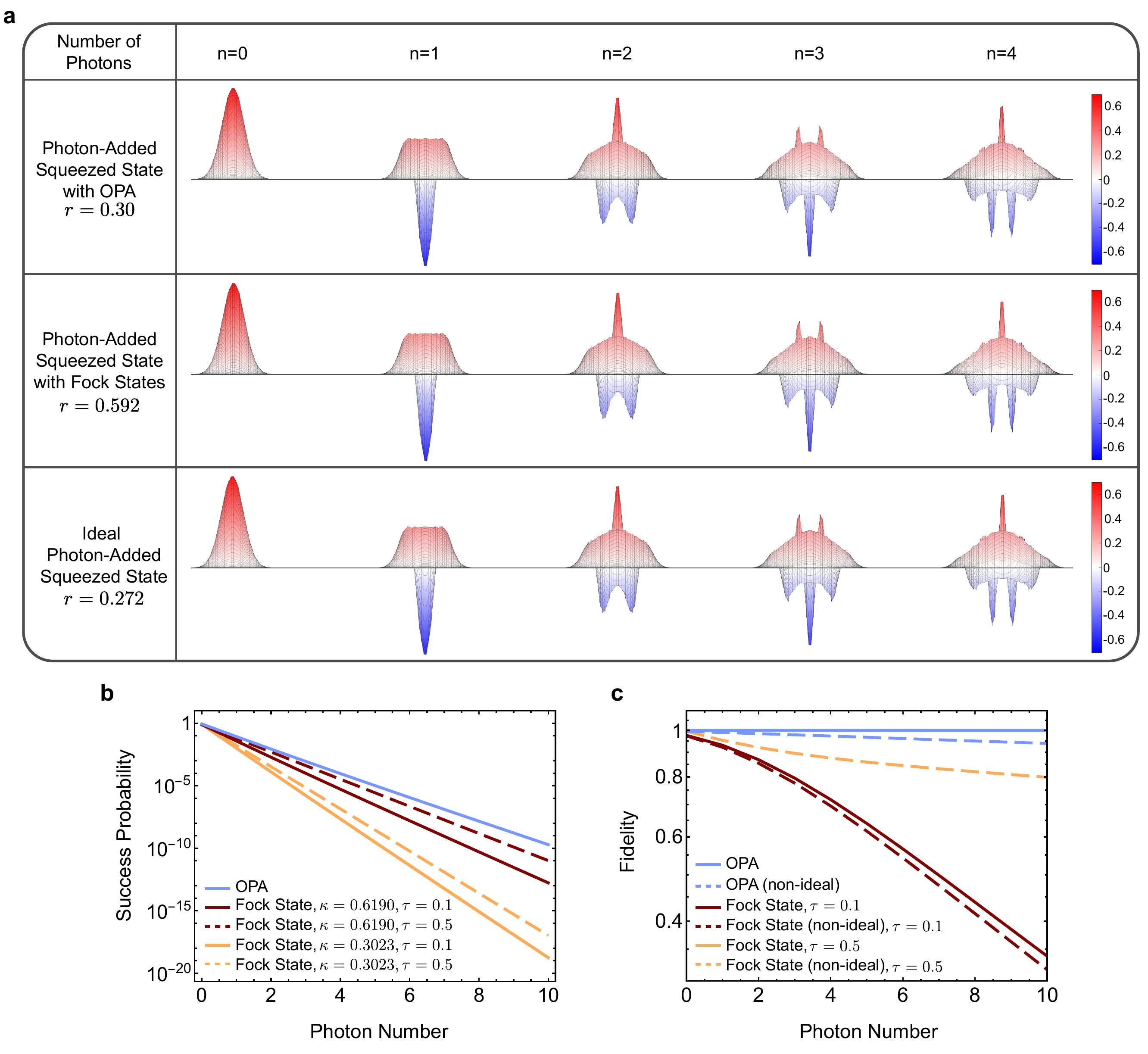}\hspace*{-0.1cm}
\caption{\label{fig:figure2}\textbf{Comparison of photon-addition schemes to squeezed states.} \textbf{(a)} Wigner functions of photon-added squeezed states generated using the OPA (top row), discrete Fock-state addition (middle row), and the ideal photon-added states obtained by applying the creation operator, $\hat{a}^\dagger$ to the squeezed vacuum (bottom row), for photon numbers $n = 0$--$4$. A squeezed vacuum with $r = 0.3$ (2.61 dB) is passed through an OPA with gain $\kappa = 0.6190$, and conditioning on $n$ detected photons produces the photon-added state. The comparable Fock-addition scheme requires $r = 0.592$~(5.14 dB). For reference, the ideal photon-added squeezed state is shown for $r = 0.272$~(2.36 dB). \textbf{(b)} Success probability as a function of photon number for both schemes. The OPA scheme uses a squeezed input state with $r = 0.30$ and gain parameter \ozlem{$\kappa = 0.6190$}, shown in solid blue line. For the Fock-state scheme, the red solid and dashed lines show success probabilities for Fock states generated with $\kappa = 0.6190$ and mixed with the squeezed state at $\tau = 0.5$ and $\tau = 0.1$, respectively. The orange lines show the same configuration using Fock states generated with $\kappa = 0.3023$, mixed at $\tau = 0.5$ (solid) and $\tau = 0.1$ (dashed). \textbf{(c)} Fidelity of the generated states with respect to the ideal photon-added squeezed states, shown for ideal~(solid) and non-ideal~(dashed) detection conditions. The non-ideal case uses a detector efficiency of $\eta = 0.95$ and a dark-count rate of 20~cps, corresponding to a probability of $4 \times 10^{-7}$ (see Methods). For the OPA scheme, the fidelity is optimised against the ideal photon-added squeezed state with squeezing $r = 0.272$. The discrete Fock-addition scheme is likewise optimised, which occurs for an input squeezing of $r = 0.592$ when compared to the same ideal target state with $r = 0.272$. Non-ideal fidelities are evaluated with respect to this same ideal target. \ozlem{Note that lines are shown as continuous lines for visual clarity; photon number is discrete and data are evaluated only at integer values.}}
\end{figure*}
\begin{equation}
\label{eq:OPA_unitary}
    U_{\mathrm{OPA}}(\kappa)=\exp[\frac{\kappa}{2}\big(\hat{a}_1\hat{a}_2-\hat{a}_1^\dagger\hat{a}_2^\dagger\big)],
\end{equation}
where $\kappa$ denotes the non-linear gain (or squeezing) parameter of the OPA, and $\hat{a}_1, \hat{a}_2$ and $\hat{a}_1^\dagger,\hat{a}_2^\dagger$ are the annihilation and creation operators corresponding to the two correlated modes produced by the OPA. The photon-number-resolving detection on one of the modes can be modelled by the projection operator
\begin{equation}
\label{eq:povm_pnrd}
\Pi_n(n) = \ketbra{n}{n}
\end{equation}
corresponding to the detection of $n$ photons, which heralds the non-Gaussian state in the remaining mode.

\section{\label{sec:results}Results}
\subsection{\label{sec:photon_added_squeezed_states}Photon-Added Squeezed States}
When the non-Gaussian source shown in Fig.~\ref{fig:figure1}(b) is employed to generate photon-added squeezed states, the input state $\ket{\psi_{\mathrm{in}}}$ is chosen to be a squeezed vacuum state of the form~\cite{weedbrook2012gaussian}
\begin{equation}
    \label{eq:squeezed_states}
    S(r)\ket{0}=\frac{1}{\sqrt{\cosh(r)}}\sum_{n=0}^\infty \frac{\sqrt{(2n)!}}{2^n n!}\tanh r^n\ket{2n},
\end{equation}
\ozlem{where $S(r)=\exp[r(\hat{a}^2-\hat{a}^{\dagger2})]$} denotes the single-mode squeezing operator with squeezing parameter $r$ and $\ket{0}$ represents the vacuum state. \ozlem{For the photon-added squeezed states, we compare both the OPA-based scheme and the discrete Fock-addition scheme against the ideal case, where photon addition is implemented by applying the creation operator to the squeezed vacuum, $\ket{\psi_{\mathrm{ideal}}} = (\hat{a}^\dagger)^n S(r)\ket{0}$, with $n$ denoting the number of added photons.}

\ozlem{Figure~\ref{fig:figure2}(a) compares photon-added squeezed states generated using three different methods: the OPA-based scheme (top row), discrete Fock-state addition (middle row), and the ideal photon-added squeezed states obtained by applying the creation operator (bottom row). To establish equivalence between these approaches, we first optimise the fidelity between the OPA-generated states and the ideal photon-added squeezed states over the input squeezing which is achieved when the input squeezing of the ideal photon-added states is $r = 0.272$. The discrete Fock-addition scheme is then independently optimised against the same ideal target, giving an input squeezing of $r = 0.592$. In both cases, the fidelities match to within $10^{-7}$ of unity, demonstrating that all three methods generate effectively identical non-Gaussian states when operated at their optimal parameters.}
\begin{figure*}[t]
\includegraphics[scale=0.38]{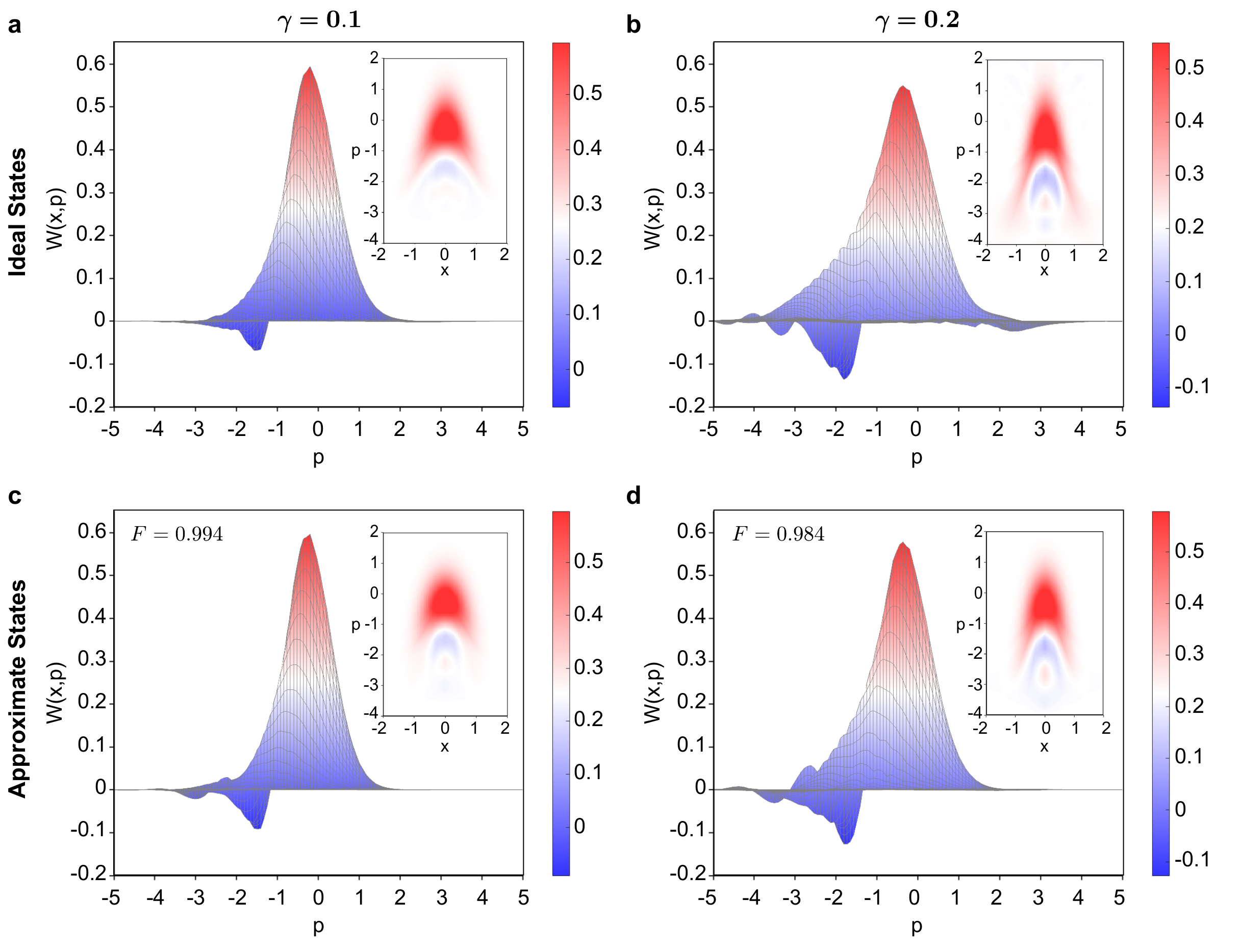}\hspace*{-0.1cm}
\caption{\label{fig:figure3}\textbf{Comparison of Wigner functions of ideal and approximate cubic-phase states.} Panels (a) and (b) depict the Wigner functions of ideal cubic-phase states with non-linearity strengths of \textbf{(a)} $\gamma=0.1$ and \textbf{(b)} $\gamma=0.2$, respectively. Panels (c) and (d) present the corresponding approximate cubic-phase states generated by seeding a coherent state into the OPA and heralding on $n=3$ photons, with OPA gain set to \ozlem{$\kappa=0.3023$}. \textbf{(c)} The fidelity between the ideal cubic-phase state and the OPA output was optimised by varying the magnitude of the input coherent state to achieve a non-linearity of $\gamma = 0.1$. The optimal match is obtained for a coherent-state amplitude of $\alpha = -1.35i$, giving a fidelity of $F = 0.994$. \textbf{(d)} For a stronger non-linearity of $\gamma = 0.2$, the optimal fidelity, $F=0.984$, is obtained for a coherent-state amplitude of $\alpha = -1.1i$. Insets
give top view of Wigner functions.
}
\end{figure*}

Figure~\ref{fig:figure2}(b) compares the success probabilities of the two schemes, accounting for the generation rate of the input Fock states. Fock states can be produced by the same Hamiltonian described in Eq.~\eqref{eq:OPA_unitary}, where a vacuum state is seeded into a spontaneous parametric down-conversion (SPDC) or weak OPA process, and one of the output modes is heralded on the detection of $n$ photons, producing an $n$-photon Fock state in the other mode. Two scenarios are considered: Fock states generated by an SPDC with \ozlem{$\kappa = 0.3023$} (orange lines) and with \ozlem{$\kappa = 0.6190$} (red lines), the latter matching the gain used in the proposed OPA scheme. Note that \ozlem{$\kappa = 0.6190$} \ozlem{which lies well outside the weak-gain SPDC regime and represents a moderately strong OPA interaction.} At each photon-addition level, our proposed OPA-based scheme exhibits substantially higher success probabilities. This improvement arises naturally because generating Fock states, $\ket{n}$, as shown in Fig.~\ref{fig:figure1}(a), requires one detector, while subsequently adding these photons to a squeezed state requires another, resulting in two heralding events. In contrast, the OPA scheme performs photon addition directly on the squeezed state and therefore needs only a single detector. The advantage becomes move evident with increasing photon number. \ozlem{For example, when ten photons are added, the success probability of our scheme exceeds that of the SPDC-based approach (with $\kappa = 0.6190$) by roughly three orders of magnitude for a beamsplitter transmissivity of $\tau = 0.5$, and remains about an order of magnitude higher even at $\tau = 0.1$.} \ozlem{As the SPDC source used to generate the Fock states is operated at relatively low gain ($\kappa = 0.3023$)}, this advantage grows dramatically, reaching nearly a $10^9$-fold improvement for ten-photon heralding. This demonstrates the exponential scaling of the success probability advantage with increasing photon number.
\begin{table*}[t]
\centering
\caption{\textbf{Comparison of cubic-phase state generation using two different schemes.} 
The left block shows states produced using the OPA-based protocol, where the OPA is seeded with a coherent state as in Fig.~\ref{fig:figure1}(b). The right block shows the coherent-state scheme in which a coherent state is interfered with a Fock state $\ket{n}$ on a balanced beamsplitter, with heralding on the $0$-photon outcome. In both cases, the OPA gain and the cavity gain used to generate the Fock states are fixed at $\kappa = 0.3023$. Notation: Opt. = Optimal, Sqz.~Corr. = Squeezing correction, Disp.~Corr. = Displacement Correction, Succ.~Prob. = Success probability.}
\begin{tabular}{|c|c|ccccc|ccccc|}
\hline
\multicolumn{2}{|c|}{\textbf{Common Parameters}} &
\multicolumn{5}{c|}{\textbf{OPA Scheme}} &
\multicolumn{5}{c|}{\textbf{Coherent-State with Fock-State Scheme}} \\
\hline
\makecell{\textbf{Non-} \\ \textbf{Linearity} \\ $\mathbf{\gamma}$} &
\makecell{\textbf{Number of} \\ \textbf{Detected} \\ \textbf{Photons}} &
\makecell{\textbf{Opt.} \\ $\mathbf{\alpha_{\mathrm{OPA}}}$} &
\makecell{\textbf{Sqz.} \\ \textbf{Corr.}} &
\makecell{\textbf{Disp.} \\ \textbf{Corr.}} &
\makecell{\textbf{Succ.} \\ \textbf{Prob.}} &
\makecell{\textbf{Fidelity}} &
\makecell{\textbf{Opt.} \\ $\mathbf{\alpha_{\mathrm{coh}}}$} &
\makecell{\textbf{Sqz.} \\ \textbf{Corr.}} &
\makecell{\textbf{Disp.} \\ \textbf{Corr.}} &
\makecell{\textbf{Succ.} \\ \textbf{Prob.}} &
\makecell{\textbf{Fidelity}} \\
\hline
$0.1$  & $n=1$ & $-1.35 i$ & $0.32$ & $2 i$     & $5.87\times10^{-2}$ & $0.986$ & $-2.1i$ & $0.29$ & $2.15i$ & $3.89\times10^{-3}$ & $0.986$ \\
$0.1$  & $n=3$ & $-1.35 i$ & $0.46$ & $2.65 i$  & $1.29\times10^{-4}$ & $0.994$ & $-2.1i$ & $0.45$ & $2.75i$ & $2.56\times10^{-6}$ & $0.995$ \\
$0.15$ & $n=3$ & $-1.35 i$ & $0.55$ & $2.65 i$  & $1.29\times10^{-4}$ & $0.989$ & $-1.95i$ & $0.54$ & $2.7i$ & $2.76\times10^{-6}$ & $0.988$ \\
$0.15$ & $n=4$ & $-1.35 i$ & $0.60$ & $2.9 i$   & $5.25\times10^{-6}$ & $0.992$ & $-1.8i$ & $0.61$ & $2.85i$ & $5.72\times10^{-8}$ & $0.993$ \\
$0.2$  & $n=3$ & $-1.1 i$  & $0.63$ & $2.5 i$   & $7.50\times10^{-5}$  & $0.984$ & $-1.7i$ & $0.61$ & $2.6i$ & $2.94\times10^{-6}$ & $0.982$ \\
$0.2$  & $n=6$ & $-1.1 i$  & $0.77$ & $3.15 i$  & $3.14\times10^{-9}$ & $0.992$ & $-1.5i$ & $0.76$ & $3.15i$ & $1.49\times10^{-11}$ & $0.992$ \\
\hline
\end{tabular}
\label{tab:cubic_phase_table}
\end{table*}

We further compare the fidelities of the two schemes in Fig.~\ref{fig:figure2}(c). When benchmarked against the target photon-added squeezed state with $r = 0.272$ under ideal detection, both the OPA scheme and the discrete Fock-addition scheme with $\tau = 0.5$ achieve near-unit fidelity. \ozlem{Although discrete Fock addition with $\tau = 0.1$ provides higher success probabilities than the $\tau = 0.5$ case, its fidelity starts at $F = 0.93$ and rapidly decreases to $F = 0.34$ as the number of added photons increases.} When detector imperfections are introduced (efficiency of 95\% and a dark-count rate of 20~cps), the fidelities of all schemes decrease, with a more pronounced reduction for the Fock-state scheme due to its use of two additional detectors. At the ten-photon level, our scheme maintains a fidelity of $F = 0.94$, compared with approximately $F = 0.80$ \ozlem{for the Fock-state scheme with $\tau = 0.5$ and $F = 0.32$ for $\tau = 0.1$, corresponding to only a $6\%$ deviation from the ideal case, versus $20\%$ and $68\%$ for the latter two configurations.}

\subsection{\label{sec:cubic_phase_states}Approximate Cubic-Phase States}
Jeng~\textit{et al.}~\cite{jeng2024strong} demonstrated that strong cubic-phase profiles can be obtained by adding photons to strong coherent states, although this was realised only through post-processing. In contrast, we show that the addition of Fock states onto strong coherent states can be implemented directly by seeding the OPA with a coherent state, which serves as $\ket{\psi_{\mathrm{in}}}$, and heralding on the detection of $n$ photons. The coherent state in the photon-number basis is given by
\begin{equation}
    \label{eq:coherent_states}
    \ket{\alpha}=\exp(-\frac{1}{2}|\alpha|^2)\sum_{n=0}^\infty \frac{\alpha^n}{\sqrt{n!}}\ket{n},
\end{equation}
\ozlemblue{where $\alpha=(x+ip)/2$ is the complex amplitude.} Applying the non-Gaussian unitary $U_{\mathrm{cubic}} = e^{i\gamma \hat{x}^3}$ to the vacuum produces the cubic-phase state, where $\gamma$ characterises the strength of the cubic non-linearity and $\hat{x} = \hat{a} + \hat{a}^\dagger$ represents the amplitude quadrature operator.
\begin{equation}
    \label{eq:ideal_cubic_phase}
    \ket{\gamma}=e^{i\gamma \hat{x}^3}\ket{0}.
\end{equation}
The fidelity between the ideal cubic-phase state in Eq.~\eqref{eq:ideal_cubic_phase} and the OPA output is maximised for a given non-linearity $\gamma$ by varying the complex amplitude of the input coherent state. Because the OPA transformation introduces additional Gaussian operations such as displacement and squeezing, these are compensated by optimising the corresponding re-displacement and squeezing parameters to ensure a fair comparison with the ideal state (refer to the Methods for a detailed explanation). 

Results are shown in Fig.~\ref{fig:figure3}. For $\gamma = 0.1$, the OPA output reaches a fidelity of $F = 0.994$ with the ideal cubic-phase state when the input coherent amplitude is $\alpha = -1.35i$, corresponding to a detection probability of $P_d = 1.29\times10^{-4}$ for $n = 3$ photons. Similarly, for $\gamma = 0.2$, a fidelity of $F = 0.984$ is achieved with an input coherent amplitude of $\alpha = -1.1i$ and a detection probability of $P_d = 7.50\times10^{-5}$. The generated state closely reproduces the main features of the ideal cubic-phase state, with slightly smoother fringes beyond $p=0$ as seen in Fig.~\ref{fig:figure3}(d), compared to the sharper interference structure in Fig.~\ref{fig:figure3}(b). For $\gamma = 0.2$, a higher fidelity can be achieved by detecting more photons; for instance, heralding on $n = 6$ photons gives $F = 0.992$, as summarised in Table~\ref{tab:cubic_phase_table}. Remarkably, if a slight reduction in fidelity (e.g., $F = 0.986$) is acceptable, a cubic-phase state with $\gamma = 0.1$ can be realised by detecting only a single photon. This significantly reduces the experimental complexity while providing a strong cubic non-linearity with a success probability nearly two orders of magnitude higher, as shown in Table~\ref{tab:cubic_phase_table}. 

\ozlemREV{In Table~\ref{tab:cubic_phase_table}, we also compare these results with the coherent-state scheme, where a coherent state is interfered with a Fock state, $\ket{n}$, on a balanced beamsplitter and heralded on the zero-photon outcome, following the method of Fig.~\ref{fig:figure1}(a). While similar fidelities can be reached using this approach, the success probabilities are typically two orders of magnitude smaller than the OPA scheme, and achieving comparable fidelities requires coherent states with slightly larger amplitudes. This highlights the practical advantage of the OPA-based method for generating high-quality cubic-phase states. By avoiding perturbative constructions and direct Fock-state implementation, the scheme provides a simpler and more scalable route to strong cubic non-linearities. As such, it brings CV optical platforms a step closer to universal quantum computing by offering an experimentally feasible path to the essential non-Gaussian gate.} 

\subsection{\label{sec:squeezed_cats}Approximate Squeezed-Cat States}
\begin{figure*}[t]
\includegraphics[scale=0.9]{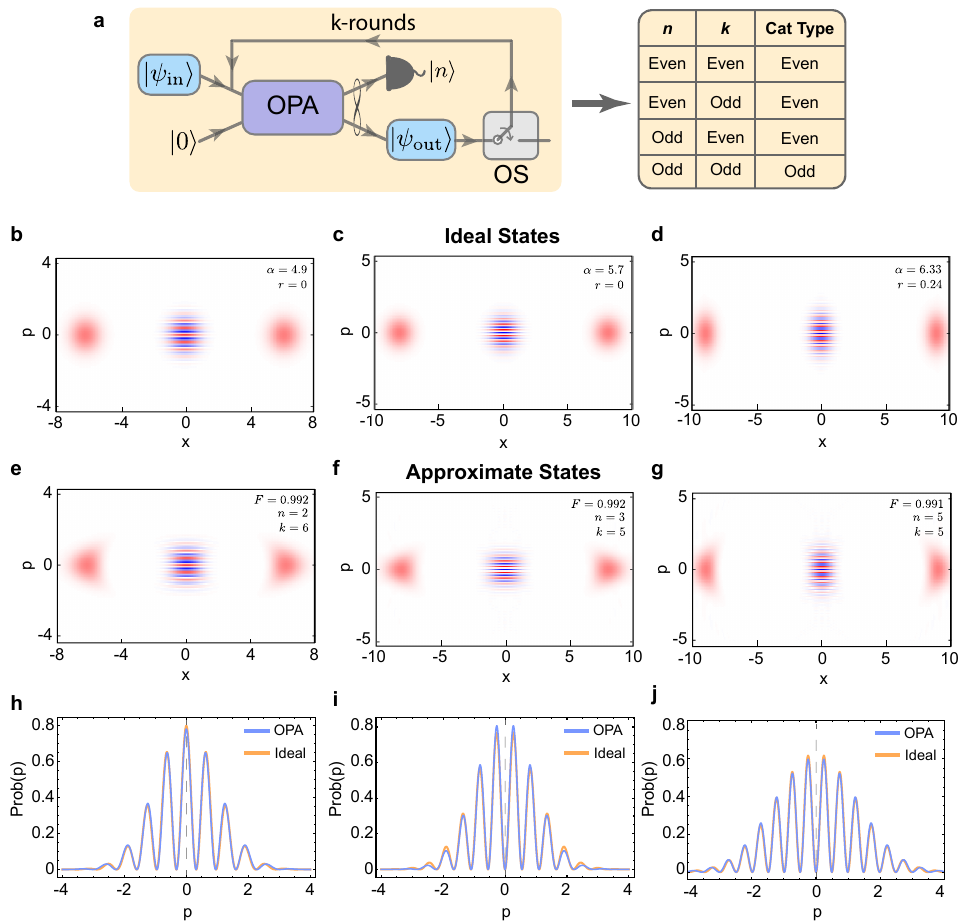}\hspace*{-0.1cm}
\caption{\label{fig:figure4}\textbf{Comparison of Wigner functions of ideal and approximate squeezed-cat states.} \textbf{(a)} OPA scheme for generating squeezed-cat states. In the first round, the OPA is seeded with a squeezed vacuum, and one of the output modes is heralded using a PNRD. Upon a successful detection event, the output passes through an optical switch~(OS) that feeds the signal back into the OPA input for the next round. If the detection fails, the OS redirects and discards the signal. After $k$ successful iterations, the OS routes the final output for further use, resulting in the desired squeezed-cat amplitude and squeezing. Panels \textbf{(b)--(d)} show the Wigner functions of the ideal squeezed-cat states with complex amplitudes $\alpha = 4.9$, $5.7$, and $6.33$, and squeezing parameters $r = 0$, $0$, and $0.24\approx2.08$~dB, respectively. Each ideal state in (b)--(d) corresponds to the approximate squeezed-cat states shown in (e)--(g), respectively. Panels \textbf{(e)--(g)} depict the Wigner functions of the approximate squeezed-cat states generated using the OPA scheme. In \textbf{(e)}, the OPA is seeded with a squeezed state of $r = -1\approx-8.69$~dB and repeated for $k = 6$ rounds, with heralding on $n = 2$ photons, giving a fidelity of $F = 0.992$ with the corresponding ideal state in (b). In \textbf{(f)}, the initial squeezing is $r = -1\approx-8.69$~dB with $k = 5$ rounds and heralding on $n = 3$ photons, giving $F = 0.992$ with the ideal state in (c). In \textbf{(g)}, the initial squeezing is $r = -0.5\approx4.34$~dB with $k = 5$ rounds and heralding on $n = 5$ photons, achieving $F = 0.991$ with the ideal state in (d). Panels \textbf{(h)--(j)} show the $p$-quadrature probability distributions of the ideal states in (b)--(d) (orange lines) and their corresponding approximate states in (e)--(g) (blue lines), respectively. Panel \textbf{(h)} corresponds to an even cat state, while panels \textbf{(i)} and \textbf{(j)} correspond to odd cat states. The OPA gain for panels (e)--(g) is set to $\kappa=0.5322$.
}
\end{figure*}

Similar to the photon-added squeezed-state protocol, the OPA is seeded with a squeezed vacuum state. Upon a successful detection, the OPA output is fed back into its input, and the process is repeated $k$ times, such that the squeezed state serves as the initial seed, while in subsequent rounds the seed becomes the OPA output from the previous iteration. The process, illustrated in Fig.~\ref{fig:figure4}(a), continues until the desired squeezing level and cat-state amplitude are achieved. For simplicity, we assume that the same photon number is heralded in each round of the protocol. The number of photons heralded per iteration, together with the total number of rounds, determines the parity of the resulting squeezed-cat state and the possible combinations are illustrated in Fig.~\ref{fig:figure4}(a). The even and odd squeezed-cat states can be written respectively as
\begin{equation}
\label{eq:cat_even}
    \ket{\psi_{\mathrm{even}}} = \mathcal{N}_{+}\big(\ket{\alpha,r} + \ket{-\alpha,r}\big),
\end{equation}
\begin{equation}
\label{eq:cat_odd}
    \ket{\psi_{\mathrm{odd}}} = \mathcal{N}_{-}\big(\ket{\alpha,r} - \ket{-\alpha,r}\big),
\end{equation}
where $\ket{\alpha,r}\equiv D(\alpha)S(r)\ket{0}$ is a displaced squeezed vacuum state in the $x-$quadrature direction. Here $\alpha$ is real and $\mathcal{N}_{\pm}$ corresponds to the normalisation constant. The displacement operator is defined as $D(\alpha)=\exp[\alpha\hat{a}^\dagger-\alpha^*\hat{a}]$ and the squeezing operator \ozlem{follows the definition given in Sec.~\ref{sec:photon_added_squeezed_states}.} Here, $r>0$ denotes squeezing, while $r<0$ denotes anti-squeezing.

The results of the protocol are shown in Fig.~\ref{fig:figure4}. Panels (b)–(d) display the Wigner functions of the ideal squeezed-cat states, constructed using Eqs.~\eqref{eq:cat_even} and~\eqref{eq:cat_odd}. The parameters $\alpha$ and $r$ are chosen to maximise the fidelities with the corresponding approximate squeezed-cat states generated via the OPA scheme, shown in panels (e)–(g), respectively. When the OPA is seeded with an anti-squeezed state ($r = -1$) and heralded on $n = 2$ photons in each round over $k = 6$ iterations, the resulting state is an even cat state, achieving a fidelity of $F = 0.992$ with respect to the ideal cat state characterised by $\alpha = 4.9$ and $r = 0$, with an overall success probability of $3.40 \times 10^{-6}$. In contrast, when an odd number of photons is heralded and the process is repeated over an odd number of rounds, odd cat and odd squeezed-cat states are obtained, as shown in panels (f) and (g), respectively. In panel (f), the OPA is seeded with an anti-squeezed state ($r = -1$), resulting in a cat state with a fidelity of $F = 0.992$ to the ideal state characterised by $\alpha = 5.7$ and $r = 0$ with an overall success probability of $5.62\times10^{-7}$. It is observed that when the input state is strongly squeezed or anti-squeezed, the output tends to form a cat state rather than a squeezed-cat state. In contrast, when a moderately squeezed or anti-squeezed seed is used, the output becomes a squeezed-cat state, as illustrated in panel (g), where the OPA is seeded with $r = -0.5$ and heralded on $n = 5$ photons over $k = 5$ rounds. This produces a squeezed-cat state with a fidelity of $F = 0.991$ relative to the ideal state ($\alpha = 6.33$, $r = 0.24$). \ozlemblue{Note that when the OPA is seeded with a squeezed state ($r > 0$), the resulting cat state is rotated by $\pi/2$ in phase space, such that the interference fringes, as shown in Fig.~\ref{fig:figure4}(h)--(j), appear along the $x$-quadrature.}

These results demonstrate that both even and odd squeezed-cat states can be efficiently engineered through the OPA-based protocol without the need for pre-generated Fock states. The high fidelities achieved across different configurations highlight that the iterative OPA process naturally reproduces the essential features of ideal squeezed cats, establishing it as a practical and scalable route for non-Gaussian state generation. The overall success probability can be further enhanced by tuning the OPA gain parameter $\kappa$, the input squeezing, and the number of rounds and detection settings, allowing the desired squeezed-cat states to be obtained at higher generation rates.

\subsection{\label{sec:gkp_breeding}GKP Breeding from Squeezed-Cats}
\begin{figure*}[!htbp]
\includegraphics[scale=0.52]{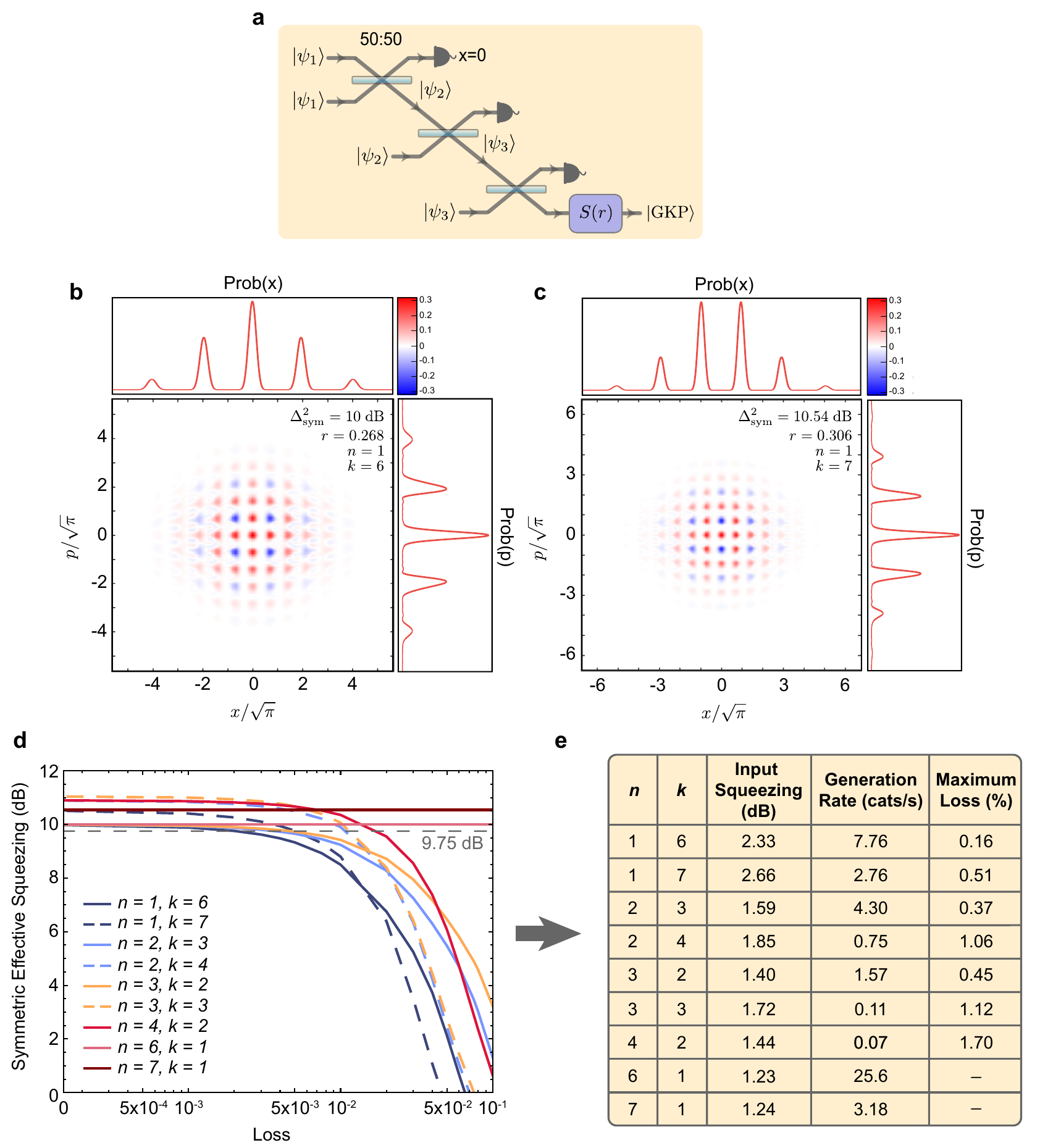}\hspace*{-0.1cm}
\caption{\label{fig:figure5}\textbf{Wigner functions of the approximate GKP states.} \textbf{(a)} Schematic of the protocol for generating approximate GKP states from squeezed cat states, following the approach of Refs.~\cite{vasconcelos2010all,weigand2018generating}. Two identical squeezed-cat states interfere on a balanced (50:50) beam splitter. One output mode is subjected to homodyne detection, and conditioned on the measurement outcome, the remaining mode is kept. A final squeezing operation is applied to adjust the lattice spacing of the resulting state to match that of an ideal GKP grid. Panels (b) and (c) show the approximate GKP states obtained from the squeezed-cat states generated using the protocol in Fig.~\ref{fig:figure4}(a). In both cases, a single-photon detection event is heralded by the PNRD, and the procedure is repeated $k=6$ and $k=7$ times, respectively, to reach the required level of squeezing in the cat state. In panel \textbf{(b)}, a logical $\ket{0_L}$ GKP state is generated from an initial squeezed state with $r = 0.268 \approx 2.33~\text{dB}$, yielding a final grid state with $10~\text{dB}$ of squeezing. Panel \textbf{(c)} shows the corresponding logical $\ket{1_L}$ state, produced from an input squeezing of $r = 0.306 \approx 2.66~\text{dB}$, resulting in a GKP state with $10.54~\text{dB}$ of squeezing. \textbf{(d)} shows the effective symmetric GKP squeezing when the optical switch in Fig.~\ref{fig:figure4}(a) is assumed to introduce loss. The dashed grey line indicates the $9.75~\text{dB}$ squeezing threshold required for fault-tolerant optical quantum computing. We compare the performance of the protocol for different numbers of PNRDs and varying numbers of OPA iterations, $k$. Note that the $n = 6, k = 1$ (in red) and $n = 7, k = 1$ (in maroon) cases require no optical switch, as no iterations are performed. These single-round schemes are therefore unaffected by switching loss and still achieve GKP squeezing of $10$~dB and $10.54$~dB from input squeezing of $r = 1.23$~dB and $r = 1.24$~dB, respectively. \textbf{(e)} summarises the protocol parameters for different combinations of $n$ and $k$, together with the corresponding squeezed-cat generation rates, assuming an optical switch operating at $50~\text{kHz}$. \ozlemREV{For the single-shot cases $k = 6$ and $k = 7$, where no optical switch is required, we instead assume detector-limited operation at $10~\text{MHz}$.} The “maximum loss’’ indicates the highest optical loss that each resulting GKP state can tolerate while still maintaining the $9.75~\text{dB}$ squeezing threshold required for fault-tolerant optical quantum computing. The OPA gain used for the generation of all GKP states is fixed at \ozlem{$\kappa = 0.7082$}.}
\end{figure*}
The squeezed-cat states produced via the OPA can serve as inputs for GKP-state generation using GKP-breeding protocols~\cite{vasconcelos2010all,weigand2018generating}, as illustrated in Fig.~\ref{fig:figure5}(a). While the squeezed cats generated in Sec.~\ref{sec:squeezed_cats} employ relatively high anti-squeezing ($\approx -8.69$~dB), we show here that by using only low to moderate levels of squeezing, GKP states with $\approx 9.75$~dB of squeezing, sufficient for fault-tolerant quantum computation~\cite{larsen2025integrated}, can be produced with single-photon detection and comparatively high success probabilities.

Ideal GKP states require infinite squeezing and are therefore unphysical in practice~\cite{gottesman2001encoding}. In contrast, approximate GKP states with finite squeezing are constructed by applying a Gaussian envelope to a lattice of displaced squeezed states in phase space. \ozlem{In the position basis, they can be written as~\cite{gottesman2001encoding}}
\ozlemblue{
\begin{equation}
\label{eq:logical_0_GKP}
    \ket{0_L}\!=\mathcal{N}_0\!\!\sum_{s\in\mathbb{Z}}\!\exp\!\bigg(\!\!\frac{-(2s\beta\Delta)^2}{2}\!\bigg)D\big({2s\beta}\big)S(-\ln\Delta)\!\ket{0},
\end{equation}
\begin{equation}
\label{eq:logical_1_GKP}    \ket{1_L}\!\!=\!\mathcal{N}_1\!\!\sum_{s\in\mathbb{Z}}\!\!\exp\!\!\bigg(\!\!\frac{-\big((2s+1)\beta\Delta\big)^2}{2}\!\bigg)\!D\big(\!(2s+1)\beta\big)\!S(\!-\!\ln\!\Delta)\!\ket{0}\!,
\end{equation}
where $\beta=\sqrt{\pi}/2$ and $\mathcal{N}_0$ and $\mathcal{N}_1$ are the normalisation constants. $\Delta \in (0,1]$ characterises the amount of squeezing in the GKP state, corresponding to $-10\log_{10}(\Delta^2)$ dB. In the limit $\Delta \rightarrow 0$, the state converges to an ideal (infinitely squeezed) GKP state.
}

To generate GKP states within our protocol, we first produce squeezed-cat states using the iterative process shown in Fig.~\ref{fig:figure4}(a). These states are then converted into GKP states via the breeding protocol of Refs.~\cite{vasconcelos2010all,weigand2018generating}, illustrated in Fig.~\ref{fig:figure5}(a). In this approach, two identical squeezed-cat states are interfered on a 50:50 beamsplitter, and one output mode is measured via homodyne detection. For clarity of presentation, we post-select on outcomes close to $x=0$ to directly demonstrate that our squeezed-cat states can form GKP grid states. However, in a realistic implementation, discarding all other outcomes would unnecessarily reduce the overall success probability. As shown in Ref.~\cite{weigand2018generating}, GKP states can be produced for arbitrary homodyne outcomes, provided that suitable Gaussian corrections such as displacement and (anti-)squeezing are applied. In our scheme, a final squeezing or anti-squeezing operation is sufficient to realign the quadratures and set the correct GKP lattice spacing. 

The resulting GKP states are shown in Fig.~\ref{fig:figure5}(b) and (c). In these examples, the squeezed-cat states are generated from initial squeezed vacuum states with squeezing parameters $r = 0.268 \approx 2.33$~dB and $r = 0.306 \approx 2.66$~dB, respectively. The output of the OPA is fed back iteratively, as described in Fig.~\ref{fig:figure4}(a), to grow the squeezed-cat states. For panels (b) and (c), this process is repeated $k = 6$ and $k = 7$ times, respectively, with a single-photon detection event heralded at each iteration. In these examples, we assume an ideal, lossless optical switch. Remarkably, using only low levels of input squeezing (below 3 dB) and single-photon detections in each iteration, our protocol generates GKP states with effective squeezing above the fault-tolerance threshold of 9.75 dB, such as 10 dB and 10.54 dB in Fig.~\ref{fig:figure5}(b) and (c), respectively. 

For each configuration, a larger number of iterations $k$ enables the protocol to tolerate higher optical loss, as additional rounds generate a more strongly squeezed-cat state. Moreover, for every combination of $(n, k)$, the squeezing parameter $r$ of the initial squeezed-vacuum state is optimised to maximise the effective GKP squeezing in the lossless case. The corresponding values of $r$ (in dB) for each configuration are listed in Fig.~\ref{fig:figure5}(e). For instance, the configuration with $n = 1$ and $k = 7$ can tolerate up to $0.51\%$ optical loss, compared to only $0.16\%$ for $n = 1$ and $k = 6$. However, this increased robustness comes at the cost of a lower generation rate: $2.76$ cats/s for $k = 7$ versus $7.76$ cats/s for $k = 6$. The generation rate is calculated by assuming an optical switch operating at 50 kHz and multiplying this by the overall success probability of the squeezed-cat state for each $(n,k)$ configuration. 

Among all configurations, the highest tolerance to optical loss is achieved for $n = 4$, $k = 2$, which can withstand up to $1.70\%$ loss. The second-best performance is obtained for $n = 3$, $k = 3$, with a tolerance of $1.12\%$ loss; however, this configuration offers a slightly higher squeezed-cat generation rate, making it more favourable in practice. For reference, state-of-the-art optical switches typically introduce around $0.2$ dB of loss, corresponding to approximately $4.5\%$, which already exceeds the tolerable loss per round for all configurations considered. 
\ozlem{It is important to note that the loss thresholds reported in Fig.~\ref{fig:figure5}(d) refer to the allowable loss per round. For protocols involving multiple rounds ($k>1$), the total tolerated transmission scales as $\eta^{k}$, where $\eta$ is the per-round transmissivity.}
\begin{figure*}[!htbp]
\includegraphics[scale=0.54]{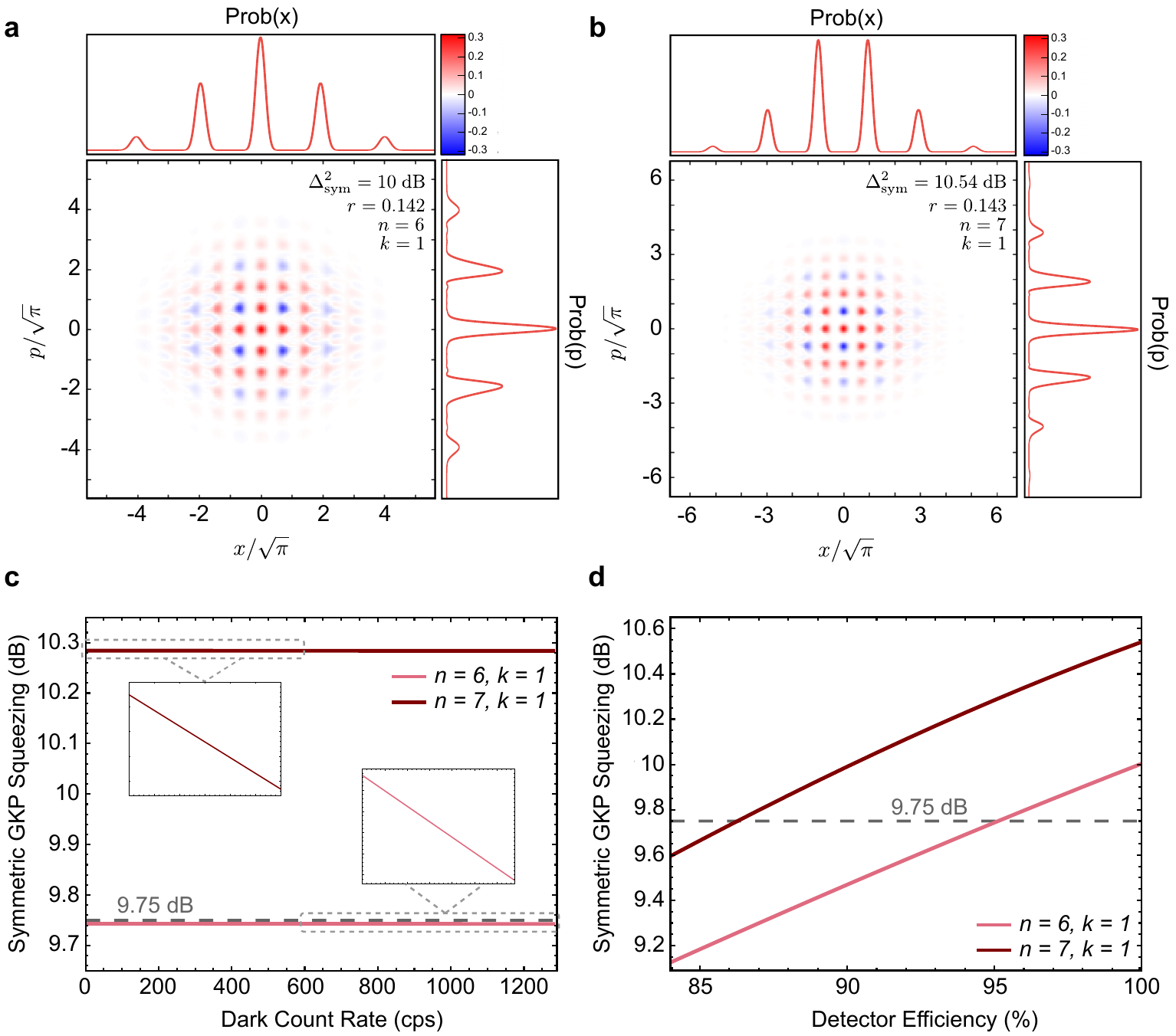}\hspace*{-0.1cm}
\caption{\label{fig:figure6}\textbf{GKP-state generation without optical-switch loss for single-round squeezed-state production with $\mathbf{n=6}$ and $\mathbf{n=7}$ photon detections.} \textbf{(a)} A logical $\ket{0_L}$ GKP state is generated from an initial squeezed state with $r=0.142\approx1.23~\mathrm{dB}$, producing a grid state with $10~\mathrm{dB}$ of squeezing when heralding on $n=6$ photons during squeezed-cat preparation. \textbf{(b)} shows the corresponding logical $\ket{1_L}$ state, produced from an input squeezing of $r=0.143\approx1.24~\mathrm{dB}$ by heralding on $n=7$ photons during squeezed-cat generation, giving a GKP state with $10.54~\mathrm{dB}$ of squeezing. \textbf{(c)} shows the performance of the GKP states when the idler is heralded using inefficient detectors during squeezed-cat generation. The detector efficiency is fixed at $95\%$, while the dark-count rate (cps) is varied, assuming a detection window of $1~\mathrm{ns}$. \textbf{(d)} illustrates the performance of the GKP states when the idler detector's dark-count rate is fixed at $20~\mathrm{cps}$ while the detector efficiency is varied. Although the dark-count rate is held constant, it is coupled to the efficiency, so the effective dark-count rate also changes (see Methods for details). The dashed grey line in panels (c) and (d) indicates the fault-tolerance threshold, and the OPA gain in both plots is set to $\kappa = 0.7082$.}
\end{figure*}

\ozlem{While the performance of our protocol could further benefit from advances in optical switching technologies, this challenge is not fundamental to the scheme itself. Rather, it reflects the current state of experimental hardware and is shared by all iterative squeezed-cat generation protocols~\cite{winnel2024deterministic}. Fiber-based optical switches offer low loss but are limited in speed, whereas free-space implementations can reach MHz rates~\cite{sonoyama2023non} at the expense of higher loss. However, in our protocol, the optical switch is only required when the process is iterated. If the OPA output is not fed back for further iterations and a higher photon number is heralded in a single round (e.g., $n = 6$ or $7$), a sufficiently strong squeezed-cat state can still be generated without the need for any optical switching, albeit with a lower success probability. For example, a 10 dB GKP state can be generated from an input squeezed state of $1.23$~dB by heralding on $n = 6$, with a success probability of $P_d = 2.59 \times 10^{-6}$ as shown in Fig.~\ref{fig:figure6}(a). Similarly, heralding on $n = 7$ with an input squeezing of $1.24$~dB yields a $10.54$~dB GKP state, with a success probability of $P_d = 3.18 \times 10^{-7}$ as shown in Fig.~\ref{fig:figure5}(b). It is worth noting that the success probability can be substantially increased by tuning the OPA gain to \ozlem{$\kappa = 0.9694$}. For example, with $n = 7$ and an input state squeezed by $1.38$~dB, the same GKP state can be generated with an improved success probability of $P_d = 1.44 \times 10^{-5}$. Without the optical switch, the experiment becomes limited by the detector repetition rate. Assuming a realistic detector rate of $10$~MHz, the squeezed-cat generation rate increases significantly, from $3.18$~cats/s to approximately $144.45$~cats/s.}

\ozlem{Furthermore, Figs.~\ref{fig:figure6}(c) and (d) shows the performance of the GKP states generated by heralding on $n=6$ and $n=7$ photons when detector imperfections are taken into account. In panel (c), the detector efficiency of the idler mode is fixed at $95\%$ while the dark-count rate is varied. Although the $n=6$ state approaches the fault-tolerance threshold, it drops slightly below it for all dark-count values considered. In contrast, the $n=7$ state remains above the threshold across the entire range of dark-count rates. Panel (d) presents the complementary scenario, where the dark-count rate is fixed at $20~\mathrm{cps}$ and the detector efficiency is varied. In this case, the generated GKP states are noticeably more sensitive to detector efficiency than to dark counts: the effective squeezing increases with higher efficiency, rather than remaining approximately constant. The $n=6$ state
exceeds the fault-tolerance threshold only for efficiencies above approximately $95.1\%$, whereas the $n=7$ state is substantially more robust, remaining above threshold for efficiencies exceeding $86\%$. This behaviour is expected, as the ideal $n=7$ state achieves a higher intrinsic squeezing level (10.54 dB) compared to the $n=6$ case (10 dB). For this reason, the $n=7$ scheme is more suitable for GKP-state generation. Both approaches offer a viable alternative to iterative squeezed-cat breeding, enabling the production of GKP states above the fault-tolerance threshold.}

Beyond detector imperfections, the remaining resources required by our protocol
are already experimentally accessible. The squeezing levels we employ are below 3 dB, which are readily achievable in both free-space and guided-wave platforms. For reference, free-space optical systems have demonstrated squeezing as high as 12.5–15 dB~\cite{vahlbruch2016detection,shajilal202212}, while fiber-based and fully guided-wave implementations have achieved up to 7–8 dB~\cite{ast2013high, liu2024entanglement}. Furthermore, our protocol requires only standard detection resources: either single-photon detection combined with multiple iterations, or a single PNRD event heralding a higher photon number (e.g., $n = 6$ or $7$), which achieves the same GKP squeezing without any iterative steps. This is in contrast to other GKP state-generation schemes that rely on significantly higher squeezing levels and multiple high-photon-number heralding events. For example, the protocol by Takase \textit{et al.}~\cite{takase2023gottesman} requires four squeezed states with $10.1$–$15.1$~dB of squeezing injected into a GBS-like circuit, and three PNRDs that simultaneously detect $n = 14$ photons to achieve a 10~dB GKP state. A related approach by Winnel \textit{et al.}~\cite{winnel2024deterministic} also generates GKP states by first producing squeezed-cat states and then applying a GKP breeding protocol~\cite{vasconcelos2010all,weigand2018generating}. However, their method requires access to higher photon-number Fock states (e.g. $n = 10$) and inline squeezing on the order of 6~dB to reach the required grid-state quality deterministically. Recently, Hanamura \textit{et al.}~\cite{hanamura2025beyond} introduced an optimisation framework for general multimode non-Gaussian state generators, showing how non-Gaussian control parameters can be tuned to improve photon-number requirements and success probabilities for a wide range of target states, including cat, cubic-phase, and GKP states. Their work provides valuable architectural insight into how non-Gaussian resources may be engineered more efficiently. In a complementary direction, our approach pursues the same goal but takes a further step toward experimental realisation by providing a unified optical platform that physically generates these states using only low levels of squeezing, heralded detection and avoiding high-photon Fock states.

\section{\label{sec:discussion}Discussion}
In this work, we introduced a unified and experimentally feasible optical framework for generating a broad class of non-Gaussian resource states that underpin quantum technologies. Using a single optical platform combining coherent or squeezed-vacuum inputs, parametric amplification and heralded photon detection, we generate squeezed-cat states, photon-added states, cubic-phase-like states and fault-tolerant GKP grid states. This shows that these states, which are often treated as independent resources for computing, communication, or metrology, can in fact be produced within one coherent platform.

A key advantage of our approach is that it operates in a low-squeezing regime ($<3$~dB) and only requires Gaussian operations combined with single-photon or low-number PNRD measurements. Despite its simplicity, the protocol achieves GKP squeezing above the 9.75 dB fault-tolerance threshold, while also allowing access to other non-Gaussian states relevant for quantum repeaters, phase estimation, bosonic error correction, and CV gate implementation. In contrast to alternative schemes that rely on large Fock states, high inline squeezing or multiple simultaneous multi-photon heralding events, our protocol reduces experimental overhead while remaining compatible with fiber, free-space, and integrated photonics. The principal limitation of our protocol comes from optical loss introduced by the optical switch when multiple iterations are required to generate the target state. However, this constraint is technological rather than fundamental, and can be alleviated through switch-free single-shot heralding, \ozlemREV{spatial multiplexing}, or integrated feedforward implementations.

Overall, our results provide a pathway towards a versatile, modular source of non-Gaussian states for optical quantum information. By unifying cat-state generation, photon addition, cubic-phase resource production and GKP encoding within a single experimental framework, this work forms a foundation for scalable optical platforms spanning quantum computing, secure communications, and precision metrology.

\section{\label{sec:methods}Methods}
\subsection{\label{sec:theory_of_photon_added_states}Success Probabilities and Fidelities of the Photon-Added Squeezed States}
Figure \ref{fig:figure1}(a) illustrates the standard procedure for generating photon-added squeezed states, in which a squeezed state of the form in Eq.~\eqref{eq:squeezed_states} is interfered with an $\ketbra{n}{n}$ Fock state and the output is heralded by projecting onto vacuum. The required Fock states are themselves produced by seeding an SPDC source with vacuum and heralding on the detection of $n$-photons where the unnormalised state can be expressed as
\begin{align}
    \label{eq:fock_state_prep}
    \rho_{\text{Fock}}(n,\kappa)=&\text{Tr}_2\big[\big(\Pi_n(n)\otimes\mathbb{I}_d\big)U_{\text{OPA}}(\kappa)\ketbra{0}{0}^{\otimes 2}\\ \nonumber
    &U_{\text{OPA}}(\kappa)^\dagger\big(\Pi_n(n)\otimes\mathbb{I}_d\big)^\dagger\big].
\end{align}
Here the SPDC source is governed by the same unitary as the OPA, $U_{\text{OPA}}(\kappa)$, defined in Eq.~\eqref{eq:OPA_unitary}. Heralding is implemented via the projector 
$\Pi_n(n)$ in Eq.~\eqref{eq:povm_pnrd}, and $\mathrm{Tr}_{2}$ denotes the 
partial trace over the measured mode. $\mathbb{I}_d$ is the identity operator on a $d$-dimensional Hilbert space. The success probability for generating the Fock-state is $P_{\text{Fock}}(n,\kappa) = \mathrm{Tr}[\rho_{\text{Fock}}(n,\kappa)]$, and the corresponding normalised state is $\tilde{\rho}_{\text{Fock}}(n,\kappa) = \rho_{\text{Fock}}(n,\kappa) / P_{\text{Fock}}(n,\kappa)$. The generated state then can be used to produce photon added squeezed states. In this case, the unnormalised output state of Fig.~\ref{fig:figure1}(a) can be derived as
\begin{align}
    \label{eq:ideal_squeezed_state}
    \rho_n(r,n,\kappa)\!=\!& \text{Tr}_2\big[\big(\mathbb{I}_d\!\otimes\!\ketbra{0}{0}\!\big)\text{BS}(1/2)\big(\rho_{\text{Sqz}}(r)\!\otimes\!\tilde{\rho}_{\text{Fock}}(n,\kappa)\big)\nonumber \\
    & \text{BS}(1/2)^\dagger\big(\mathbb{I}_d\!\otimes\!\ketbra{0}{0}\!\big)^\dagger\big],
\end{align}
where $\rho_{\text{Sqz}}(r) = S(r)\ketbra{0}{0}S^\dagger(r)$, with $S(r)\ket{0}$ being the squeezed-vacuum state defined in Eq.~\eqref{eq:squeezed_states}. The operator $\text{BS(1/2)}$ denotes a beamsplitter with transmissivity $\tau = 1/2$, expressed as
\begin{equation}
\label{eq:beamsplitter}
\text{BS}(\tau)=\mathrm{exp}[\mathrm{cos}^{-1}({\sqrt{\tau}})(\hat{a}^\dagger\hat{b}-\hat{a}\hat{b}^\dagger)].
\end{equation}
Here \text{$\hat{a}$} and \text{$\hat{b}$} are the annihilation operators, while \text{$\hat{a}^\dagger$} and \text{$\hat{b}^\dagger$} are the creation operators of the two modes, respectively. The success probability of this state is given by 
$P_n(r,n,\kappa) = \mathrm{Tr}[\rho_n(r,n,\kappa)]$, and the corresponding normalised photon-added squeezed state is $\tilde{\rho}_n(r,n,\kappa) = \rho_n(r,n,\kappa) / P_n(r,n,\kappa)$. Since the overall probability must also include the heralded generation of the Fock state, the total success probability is $P_T(r,n,\kappa) = P_{\text{Fock}}(n,\kappa)\, P_n(r,n,\kappa)$, which is used to generate the curves in Fig.~\ref{fig:figure2}(b).

In contrast, the photon-added squeezed states generated by the OPA scheme in Fig.~\ref{fig:figure1}(b) are obtained from
\begin{align}
    \label{eq:squeezed_OPA_photon_states}
    \rho_{\text{OPA}}(r,n,\kappa)\!=\!\text{Tr}_2\!\big[\!&\big(\mathbb{I}_d\otimes\Pi_n(n)\big)U_{\text{OPA}}(\kappa)\big(\rho_{\text{Sqz}}(r)\otimes\ketbra{0}{0}\!\big)\nonumber \\
    &U_{\text{OPA}}^\dagger(\kappa)\big(\mathbb{I}_d\otimes\Pi_n(n)\big)^\dagger\big],
\end{align}
where the success probability of the process is $P_d(r,n,\kappa) = \mathrm{Tr}[\rho_{\text{OPA}}(r,n,\kappa)]$, which is used to generate the data in Fig.~\ref{fig:figure2}(b). The corresponding normalised state is $\tilde{\rho}_{\text{OPA}}(r,n,\kappa) = \rho_{\text{OPA}}(r,n,\kappa) / P_d(r,n,\kappa)$. 

Figure~\ref{fig:figure2}(c) illustrates the fidelities of the output of the OPA scheme as well the photon-added squeezed states using Fock-states. When the detector is ideal, the OPA output states are compared with the ideal photon-added states in Eq.~\eqref{eq:ideal_squeezed_state}, and the fidelity between the two states is computed as
\begin{equation}
    \label{eq:fidelity}
    F(\rho,\sigma)=\text{Tr}\bigg[\sqrt{\sqrt{\rho}\sigma\sqrt{\rho}}\bigg]^2.
\end{equation}
It is important to note that the OPA output is further squeezed relative to the initial input state. To allow a fair comparison between the OPA output $\tilde{\rho}_{\text{OPA}}(r_1,n,\kappa)$ and the ideal photon-added squeezed state $\tilde{\rho}_{n}(r_2,n,\kappa)$, the input squeezing parameter of the ideal scheme must be optimised. This optimisation is defined by
\begin{equation}
    \label{eq:fidelity_optimisation_photon_added_states}
    F_\text{Sqz}(n,\kappa)=\max_{r_2}F(\Tilde{\rho}_{\text{OPA}}(r_1,n,\kappa),\Tilde{\rho}_{n}(r_2,n,\kappa)).
\end{equation}

Figure~\ref{fig:figure2}(c) also shows the reduction in fidelity when detector inefficiencies and dark counts are included. These effects are modelled using the method described in Ref.~\cite{erkilicc2023surpassing}, where the detector is treated as a thermal-loss channel. The detector efficiency $\eta$ is modelled as a beamsplitter with transmittivity $\eta$, and dark counts are incorporated by mixing a thermal state into the same beamsplitter. In this case, let the unheralded OPA output be $\rho'_{\text{OPA}}$. The effect of the detector on the idler arm prior to measurement can then be written as
\begin{equation}
    \rho_\text{OPA}''\!=\!\label{eq:dark_count_opa}
    \text{Tr}_3\big[\big(\mathbb{I}_2\otimes\text{BS}(\eta)\big)\big(\rho'_{\text{OPA}}\!\otimes\!\rho_{\text{th}}(\bar{n})\big)\big(\mathbb{I}_2\otimes\text{BS}(\eta)\big)^\dagger\big],
\end{equation}
where $\rho_{\text{th}}(\bar{n})$ is a thermal state with mean photon number $\bar{n}$, used to model dark counts and is defined as
\begin{equation}
    \label{eq:thermal_state}
    \rho_\text{th}=\sum_{n=0}^\infty \frac{\bar{n}^n}{(1+\bar{n})^{n+1}}\ketbra{n}{n}.
\end{equation}
Here, the dark-count rate is coupled to the detector efficiency according to
\begin{equation}
\label{eq:mean_photon_number_dark}
    R_d\times D_w = \frac{(1-\eta)\bar{n}}{1+(1-\eta)\bar{n}},
\end{equation}
where $R_d$ denotes the dark-count rate in counts per second~(cps) and $D_w$ is the detection window (typically given in ns). Note that both the OPA-generated states and the photon-added squeezed states based on Fock-state addition are compared with the ideal states defined in Eq.~\eqref{eq:ideal_squeezed_state}. In all cases, the comparison is made against the ideal state with the optimised value of $r_2$ prior to introducing detector loss and noise.

\subsection{\label{sec:cubic_phase_methods}Success Probabilities and Fidelities of the Cubic-Phase States}
The approximate cubic-phase states described in Sec.~\ref{sec:cubic_phase_states} can be generated within the OPA scheme by seeding the OPA with a coherent state. The unnormalised output of the OPA is given by
\begin{align}
    \label{eq:approximate_cubic_phase_state}
    \rho_\text{Cubic}(\alpha,n,\kappa)\!=\!\text{Tr}_2[&\big(\mathbb{I}_d\otimes\Pi_n(n)\big)U_{\text{OPA}}(\kappa)\big(\!\ketbra{\alpha}{\alpha}\otimes\ketbra{0}{0}\!\big)\nonumber \\
    &U_{\text{OPA}}^\dagger(\kappa)\big(\mathbb{I}_d\otimes\Pi_n(n)\big)^\dagger],
\end{align}
where $\ket{\alpha}$ is the input coherent state defined in Eq.~\eqref{eq:coherent_states}. The success probability of the process is $P_d(\alpha,n,\kappa)=\text{Tr}[\rho_\text{Cubic}(\alpha,n,\kappa)]$ and the corresponding normalised state is $\Tilde{\rho}_\text{Cubic}(\alpha,n,\kappa)= \rho_\text{Cubic}(\alpha,n,\kappa)/P_d(\alpha,n,\kappa)$.

The OPA output includes additional Gaussian operations, such as squeezing and displacement. Therefore, before evaluating the fidelity between the OPA output and the ideal cubic-phase state $\ket{\gamma}$ in Eq.~\eqref{eq:ideal_cubic_phase}, these Gaussian contributions must be corrected, which is achieved by applying a displacement followed by a single-mode squeezing operation
\begin{align}
    \label{eq:corrections_OPA_cubic_phase}
    \rho_c(r,\alpha,\beta,n,\kappa)=S(r)D(\beta)\Tilde{\rho}_\text{Cubic}(\alpha,n,\kappa)D^\dagger(\beta)S^\dagger(r),
\end{align}
where $D(\beta)$ is the displacement operator defined in Sec.~\ref{sec:squeezed_cats}, and $S(r)$ is the single-mode squeezing operator defined in Sec.~\ref{sec:photon_added_squeezed_states}. Therefore, to maximise the fidelity with a cubic-phase state of a given non-linearity $\gamma$, the parameters $r$, $\alpha$, and $\beta$ are optimised according to
\begin{equation}
    \label{eq:max_fidelity_cubic}
    F_{\text{Cubic}}(n,\kappa,\gamma)=\max_{r,\alpha,\beta}F(\rho_c(r,\alpha,\beta,n,\kappa),\ketbra{\gamma}{\gamma}).
\end{equation}

\ozlemREV{In Table~\ref{tab:cubic_phase_table}, we also compare the ideal cubic-phase states with those generated using the scheme of Fig.~\ref{fig:figure1}(a), where a coherent state is interfered with a Fock state $\ket{n}$ on a balanced beamsplitter and heralded on the zero-photon outcome. The output state of this scheme can be expressed as
\begin{align}
    \label{eq:coherent_fock_cubic}
    \rho_{\text{F-Cubic}}(\alpha,\!n,\!\kappa)\!=&\text{Tr}_2\!\big[(\mathbb{I}_d\!\otimes\!\Pi_0)\text{BS}(1/2)\big(\!\ketbra{\alpha}{\alpha}\!\otimes\!\Tilde{\rho}_{\text{Fock}}(n,\kappa)\!\big)\nonumber\\
    &\text{BS}(1/2)^\dagger(\mathbb{I}_d\!\otimes\!\Pi_0)^\dagger\big],
\end{align}
where $\Tilde{\rho}_{\text{Fock}}(n,k)$ is the normalised version of the state defined in Eq.~\eqref{eq:fock_state_prep}. The probability of heralding the zero-photon outcome is given by $P_{\text{F-Cubic}}(\alpha,n,\kappa)=\mathrm{Tr}[\rho_{\text{F-Cubic}}(\alpha,n,\kappa)]$ and the normalised output state becomes $\Tilde{\rho}_{\text{F-Cubic}}(\alpha,n,\kappa)=\rho_{\text{F-Cubic}}(\alpha,\!n,\!\kappa)/P_{\text{F-Cubic}}(\alpha,n,\kappa)$. The overall success probability for preparing the approximate cubic-phase state is therefore $P_{\text{tot-Cubic}}(\alpha,n,\kappa)=P_{\text{Fock}}(n,\kappa)\, \times P_{\text{F-Cubic}}(\alpha,n,\kappa)$.}

\ozlemREV{As in the OPA-based scheme, the state produced by the coherent-Fock scheme undergoes several additional Gaussian operations. To make a fair comparison with the ideal cubic-phase state, these operations needs to be inverted. This is achieved by applying Eq.~\eqref{eq:corrections_OPA_cubic_phase}, with $\Tilde{\rho}_{\text{Cubic}}(\alpha,n,\kappa)$ replaced by $\Tilde{\rho}_{\text{F-Cubic}}(\alpha,n,\kappa)$. Similarly, the fidelity with respect to the ideal cubic-phase state is optimised over $r$, $\alpha$, and $\beta$, as specified in Eq.~\eqref{eq:max_fidelity_cubic}.}

\subsection{\label{sec:squeezed_cat_methods}Success Probabilities and Fidelities of the Squeezed-Cat States}
The scheme for generating squeezed-cat states is shown in Fig.~\ref{fig:figure4}(a), where the OPA output is iteratively fed back into its input. In the first round ($k = 1$), the seed is a squeezed state, so the OPA output is given by Eq.~\eqref{eq:squeezed_OPA_photon_states}. In the successive rounds (i.e. $k = 2$), the unnormalised OPA output becomes
\begin{align} 
\label{eq:OPA_squeezed_cat} 
\rho_{\text{OPA}}^k(r,n,\kappa)=&\text{Tr}_2\big[\big(\mathbb{I}_d\otimes\Pi_n(n)\big)U_\text{OPA}(\kappa)\nonumber \\ 
&\big(\Tilde{\rho}_{\text{OPA}}^{k-1}(r,n,\kappa)\otimes \ketbra{0}{0}\big)U_\text{OPA}^\dagger(\kappa)\nonumber \\
&\big(\mathbb{I}_d\otimes\Pi_n(n)\big)^\dagger\ \big],
\end{align}
where $P_d^k(r,n,\kappa)$ is the success probability of the $k$-th round defined as $P_d^k(r,n,\kappa)=\text{Tr}[\rho_{\text{OPA}}^k(r,n,\kappa)]$. The corresponding normalised state is $\Tilde{\rho}_{\text{OPA}}^k(r,n,\kappa)=\rho_{\text{OPA}}^k(r,n,\kappa)/P_d^k(r,n,\kappa)$. The overall success probability for generating the squeezed-cat state after $k$ successful iterations is then
\begin{equation}
    P_{\text{tot}}(r,n,\kappa) = \prod_{j=1}^{k} P_d^j(r,n,\kappa).
\end{equation}

To ensure a fair comparison between the approximate squeezed-cat states and the ideal states defined in Eqs.~\eqref{eq:cat_even} and~\eqref{eq:cat_odd}, we optimise the parameters $\alpha$ and $r$ of the ideal cat states so as to maximise the fidelity with the generated states. This is achieved by 
\begin{equation}
    \label{eq:fidelity_squeezed_cats}
    F_{\text{Sqz-cat}}(n,\kappa)=\max_{\alpha,r_2}F\big(\Tilde{\rho}_{\text{OPA}}^k(r_1,n,\kappa),\rho_\text{ideal-cat}(\alpha,r_2)\big),
\end{equation}
where $\rho_\text{ideal-cat}(\alpha,r_2)=\mathcal{N}_\pm^2\big(\ket{\alpha,r_2}\pm\ket{-\alpha,r_2}\big)\big(\ket{\alpha,r_2}\pm\ket{-\alpha,r_2}\big)^\dagger$.

\subsection{\label{sec:loss_squeezed_cat_state_methods}Squeezed-Cat States under Optical-Switch Loss}
When the OPA output passes through an optical switch, the state inevitably experiences optical loss. We model this loss by coupling the state to vacuum via a beamsplitter with transmittivity $\tau$. In this case the lossy squeezed-cat just after the OPA for the $(k-1)$-th round can be expressed as
\begin{align} 
\label{eq:lossy_OPA} 
\rho_{\text{OPA}}^{k-1}(r,n,\kappa,\tau)\!=\!\text{Tr}_2\!\big[\text{BS}(\tau)\big(\Tilde{\rho}_{\text{OPA}}^{k-1}(r,n,\kappa)\!\otimes\!\ketbra{0}{0}\!\big)\text{BS}^\dagger(\tau)\big],
\end{align}
which is then used as the input to the OPA in the $k$-th round. Therefore, $\Tilde{\rho}_{\text{OPA}}^{k-1}(r,n,\kappa)$ in Eq.~\eqref{eq:OPA_squeezed_cat} is now replaced with $\rho_{\text{OPA}}^{k-1}(r,n,\kappa,\tau)$.

\subsection{\label{sec:GKP_methods}Inferring Squeezing Level of the GKP States}
The quality of the approximate GKP states is assessed using stabiliser 
expectation values. We consider the stabilisers $S_x = e^{i\pi \hat{x}/\alpha}$ and \ozlem{$S_p = e^{i \hat{p}\alpha}$}, where $\hat{x} = \hat{a} + \hat{a}^\dagger$ and $\hat{p} = i(\hat{a} - \hat{a}^\dagger)$. For consistency with the ideal 
GKP lattice, we set $\alpha = \sqrt{\pi}$, corresponding to the ideal grid spacing. The stabiliser expectation value is obtained by evaluating
\begin{equation}
    \label{eq:stabiliser_expectation}
    \langle S_{x(p)} \rangle=\text{Tr}[\rho_{\text{GKP}}S_{x(p)}],
\end{equation}
which is used to infer the squeezing level of the approximate GKP states~\cite{duivenvoorden2017single,larsen2025integrated} via
\begin{equation}
    \Delta^2_{x(p)}=-\frac{1}{\pi}\ln(|S_{x(p)}|^2),
\end{equation}
and the corresponding effective squeezing (in dB) is given by $-10\log_{10}\Delta^2_{x(p)}$.

\subsection{\label{sec:inefficient_detector_GKP}Modelling of Inefficient Detectors in Squeezed-Cat Preparation}
In Fig.~\ref{fig:figure6}(c) and (d), we show the effect of detector inefficiency on the symmetric GKP squeezing. As in Fig.~\ref{fig:figure2}(c), the detector is modelled using a thermal-loss channel. However, rather than introducing the thermal-loss through an additional mode, we employ the Kraus-operator formulation, as the Hilbert-space dimension required to simulate the GKP states is too large to efficiently handle a three-mode system. The Kraus operators for the thermal-loss channel can be decomposed into a pure-loss channel with transmissivity $T=\eta/G$ followed by a quantum-limited amplifier with gain $G=1+(1-\eta)\bar{n}$. The resulting Kraus-operator representation is
\begin{equation}
\label{eq:kraus_operator_thermal_channel}
    \mathcal{E}(\rho''_{\text{OPA}})=\sum_{i}^\infty\sum_k^\infty \hat{B}_k \hat{A}_i \rho'_{\text{OPA}} \hat{A}_i^\dagger \hat{B}_k^\dagger,
\end{equation}
where the operators $\hat{A}_i$ and $\hat{B}_k$ are defined respectively as
\begin{equation}
    \label{eq:pureloss_kraus}
    \hat{A}_i=\sqrt{\frac{(1-T)^i}{i!}}T^{\hat{n}/2}\hat{a}^i,
\end{equation}

\begin{equation}
    \label{eq:amplification_kraus}
    \hat{B}_i=\sqrt{\frac{1}{k!\;G}\bigg(\frac{G-1}{G}\bigg)^k}\hat{a}^{\dagger^k}G^{-\hat{n}/2}.
\end{equation}
Here $\hat{n}$ denotes the number operator, defined as $\hat{n}=\hat{a}^\dagger \hat{a}$. The summation in Eq.~\eqref{eq:kraus_operator_thermal_channel} is truncated at the Hilbert-space dimension used for the OPA density matrix.

\noindent \textbf{Data availability:}
The data that supports the findings of this study is available from the corresponding author upon reasonable request.

\noindent \textbf{Code availability:}
The codes that support the findings of this study are available from the corresponding author upon reasonable request.

\noindent \textbf{Acknowledgments:}
\ozlemREV{This research was funded by the Australian Research Council Centre of Excellence for Quantum Computation and Communication Technology (Grant No. CE170100012) and by A*STAR grants C230917010 (Emerging Technology), C230917004 (Quantum Sensing) and Q.InC Strategic Research and Translational Thrust.}

\noindent \textbf{Author contributions:}
\ozlem{SMA conceived the project and proposed the use of an optical parametric amplifier. OE developed the full protocol, including the methods for generating GKP states and other non-Gaussian states from the OPA-based architecture, with input from AD, BS and TCR. PKL, TCR and SMA supervised the project. OE performed the simulations and drafted the manuscript with contributions from all authors.}
\bibliography{apssamp}

\clearpage
\twocolumngrid 
\onecolumngrid 
\appendix
\section*{Supplementary Material}
\section*{\label{sec:GKP_parameters}Supplementary Note I: Parameters of the GKP Breeding from Squeezed-Cat States}
In this section, we summarise the parameters used to generate the squeezed states employed in the GKP-breeding protocol. The squeezed-cat states are produced using the circuit shown in Fig. 4(a) of the main text, and these states are subsequently used to generate the GKP states via the circuit in Fig. 5(a), following the methods of Refs.~\cite{vasconcelos2010all, weigand2018generating}. Supplementary Table~\ref{tab:GKP_table} lists the parameters used under the assumption of lossless optical switches and ideal detectors. When the squeezed-cat generation stage includes a lossy optical switch, the overall success probabilities and the squeezing corrections required to obtain properly aligned GKP grids change slightly.
\begin{table*}[h]
\renewcommand{\tablename}{Supplementary Table}
\centering
\caption{\textbf{Parameters used for squeezed-cat generation and GKP breeding under ideal conditions.}}
\begin{tabular}{|c|c|c|c|c|c|}
\hline
 \makecell{\textbf{Number of} \\ \textbf{Detected} \\ \textbf{Photons} $\mathbf{(n)}$} & \makecell{\textbf{Number of} \\ \textbf{Rounds} \\ $\mathbf{(k)}$} & \makecell{\textbf{OPA Gain} \\ $\mathbf{\kappa}$} & \makecell{\textbf{Input} \\ \textbf{Squeezing} \\ \textbf{(dB)}} & \makecell{\textbf{Squeezing} \\ \textbf{Correction} \\ \textbf{(dB)}} & \makecell{\textbf{Success} \\ \textbf{Probability}} \\
\hline
$1$ & $6$ & $0.7082$ & $2.33$ & $-0.54$ & $1.56\times10^{-4}$ \\
$1$ & $7$ & $0.7082$ & $2.66$ & $-1.18$ & $5.61\times10^{-5}$ \\
$2$ & $3$ & $0.7082$ & $1.59$ & $-0.54$ & $8.67\times10^{-5}$ \\
$2$ & $4$ & $0.7082$ & $1.85$ & $-1.74$ & $1.57\times10^{-5}$ \\
$3$ & $2$ & $0.7082$ & $1.40$ & $-0.52$ & $3.16\times10^{-5}$ \\
$3$ & $3$ & $0.7082$ & $1.72$ & $-2.28$ & $2.23\times10^{-6}$ \\
$4$ & $2$ & $0.7082$ & $1.44$ & $-1.74$ & $1.50\times10^{-6}$ \\
$6$ & $1$ & $0.7082$ & $1.23$ & $-0.52$ & $2.59\times10^{-6}$ \\
$6$ & $1$ & $0.9694$ & $1.37$ & $-0.52$ & $6.73\times10^{-5}$ \\
$7$ & $1$ & $0.7082$ & $1.24$ & $-1.18$ & $3.18\times10^{-7}$ \\
$7$ & $1$ & $0.9694$ & $1.38$ & $-1.19$ & $1.45\times10^{-5}$ \\
\hline
\end{tabular}
\label{tab:GKP_table}
\end{table*}

To obtain the correct GKP grid spacing, the squeezing correction shown in Fig.~5(a) is calculated as
\begin{equation}
    \label{eq:sqz_correction_GKP}
    r_{\text{corr}}=\frac{1}{4}\ln\bigg(\frac{\Delta_x^2}{\Delta_p^2}\bigg),
\end{equation}
where $\Delta_x^2$ and $\Delta_p^2$ are the quadrature variances obtained from the
expectation values of the density matrix $\rho$ as
\begin{align}
    \Delta_x^2 &= \langle \hat{x}^2 \rangle - \langle \hat{x} \rangle^2
    = \mathrm{Tr}[\rho\,\hat{x}^2] - \big(\mathrm{Tr}[\rho\,\hat{x}]\big)^2,
\end{align}
\begin{align}
    \Delta_p^2 &= \langle \hat{p}^2 \rangle - \langle \hat{p} \rangle^2
    = \mathrm{Tr}[\rho\,\hat{p}^2] - \big(\mathrm{Tr}[\rho\,\hat{p}]\big)^2,
\end{align}
with quadrature operators defined as $\hat{x}=\hat{a}+{a}^\dagger$ and $\hat{p}=i(\hat{a}-{a}^\dagger)$.

\bibliographystyle{naturemag}

%
\end{document}